\documentclass[epj]{svjour}
\usepackage{graphics}
\usepackage{amsmath}
\usepackage{amssymb}
\usepackage{amsfonts}

\usepackage{color}
\def\aeq{ \hspace{-0mm} &=& \hspace{-0mm} }
\def\slashb#1{\ \hspace{-1mm}\not\!\!#1}

\begin{document}
\title{Supersymmetric gradient flow in ${\cal N}=1$ SYM}
\author{Daisuke Kadoh\inst{1,2} 
\and Naoya Ukita\inst{3}
\thanks{\emph{Email: kadoh@keio.jp, ukita@ccs.tsukuba.ac.jp}}%
}                     
\offprints{}          
\institute{Department of Physics, Faculty of Science, Chulalongkorn University, 
Bangkok 10330, Thailand 
\and 
Research and Educational Center for Natural Sciences, Keio University, Yokohama 223-8521, Japan
\and 
Center for Computational Sciences, University of Tsukuba, Tsukuba, Ibaraki 305-8577, Japan
}
\date{Received: date / Revised version: date}
%
\abstract{
The gradient flow equation is 
derived in four-dimensional ${\cal N}=1$ supersymmetric Yang-Mills theory 
in terms of 
the component field of the Wess-Zumino gauge. 
We show that 
the flow-time derivative and supersymmetry transformation that is naively extended to 4+1 dimensions 
by replacing the four-dimensional fields with the corresponding flowed fields  commute 
with each other up to a gauge transformation. In this sense,  
the obtained flow is supersymmetric in the Wess-Zumino gauge.
We also discuss more about the symmetry of the flow equation.
\PACS{
      {PACS-key}{discribing text of that key}   \and
      {PACS-key}{discribing text of that key}
     } 
} 
\maketitle

\section{Introduction} 
\label{sec:introduction}

The gradient flow~\cite{Narayanan:2006rf,Luscher:2010iy} has been applied to various studies in lattice QCD 
\cite{Luscher:2011bx,Luscher:2013cpa,Suzuki:2013gza,Makino:2014taa,DelDebbio:2013zaa,Asakawa:2013laa,Kitazawa:2016dsl,Kitazawa:2017qab,Yanagihara:2018qqg,Taniguchi:2016ofw,Hirakida:2018uoy}.
These applications are based on 
the UV-finiteness of correlators of flowed fields and the smoothing effects obtained by the flow \cite{Luscher:2011bx}. 
It also has great potential for supersymmetric Yang-Mills (SYM) flow 
because we have not only the same applications as in QCD \cite{Bergner:2015adz,Ali:2018dnd}
but also  ones specific to SYM such as the supercurrent needed 
to construct the correct continuum limit \cite{Hieda:2017sqq,Kasai:2018koz}.

There are mainly two possibilities in defining a gradient flow equation in SYM. 
One way is to use the non SUSY flow  
as introduced in QCD~\cite{Narayanan:2006rf,Luscher:2010iy,Luscher:2013cpa}, 
while the other way is to use
a SUSY flow associated with the gradient of the SYM action.     
In the former case, the gaugino is treated as the adjoint matter coupled to the gauge field. 
In this sense, the flow is irrelevant to SUSY 
and the flowed fermions receive extra renormalizations.  
In the latter case, 
one can expect that no such extra renormalizations exist thanks to SUSY.

The SYM-gradient flow has already been given 
in terms of vector superfield~\cite{Kikuchi:2014rla}, 
which manifestly respects super and extended gauge transformations\footnote{
In the context of Langevin equation, a similar equation has been introduced by Nakazawa~\cite{Nakazawa:2003zf,Nakazawa:2003tz}}
\footnote{
The extended gauge transformation is a gauge transformation generated by a chiral superfield, 
which is defined in eq.~(\ref{ext_gauge}) in section~\ref{sec:susygf}. The ordinary gauge transformation 
$A_\mu \rightarrow A_\mu -\partial_\mu \omega$ is included in it as a partial transformation.
}.
It is, however, not straightforward to derive a flow equation in terms of component fields taking the  WZ gauge. 
In Ref.~\cite{Kikuchi:2014rla}, 
a term by which the extended gauge symmetry is fixed is introduced in the SYM-flow equation to take the Wess-Zumino gauge, 
but the obtained equation is not invariant under the ordinary gauge transformation.
It is not clear whether or not the flow is compatible with supersymmetry in the WZ gauge~\cite{deWit:1975veh,KK_Private}.

The SYM flow at the WZ gauge will be useful for the lattice calculation performed with the gauge~\cite{Bergner:2015adz,Ali:2018dnd}
and for knowing the clear difference from the non-SUSY flow. 
When we try to compare some analytic calculations performed in the component fields directly to the flow theory,  
the flow given by the  component fields will be useful. 

In this paper, we propose a natural way to derive  
a supersymmetric gradient flow equation in the Wess-Zumino gauge in ${\cal N}=1$ SYM 
and discuss the symmetry properties of the equation.   
Taking a gauge fixing term that
satisfies a condition milder than that of~\cite{Kikuchi:2014rla}, 
we are free to choose any gauge for the ordinary gauge symmetry, in other words,
the gauge symmetry is not fixed in the flow.  
We also show that the flow equation is compatible to supersymmetry 
in the sense that the commutator of the flow-time derivative and 
supersymmetry transformation that is naturally extended to $d+1$-dimensions 
vanishes up to a gauge transformation.

This paper is organized as follows.  
Starting from the definition of non-SUSY flow (the Yang-Mills flow) in section~\ref{sec:ymflow}, 
we review the SYM-gradient flow given in Ref.~\cite{Kikuchi:2014rla} in section~\ref{sec:susygf}. 
The superfield formalism is given in the Euclidean space in section~\ref{subsec:superfield} 
and the original derivation of the SYM-gradient 
flow in terms of vector superfield is shown in section~\ref{subsec:KO}. 
We discuss the issue of the original flow equation in the Wess-Zumino gauge in section~\ref{subsec:KO2}.
In section~\ref{subsec:KU}, we derive a supersymmetric gradient flow equation 
in the Wess-Zumino gauge taking a new gauge fixing term 
and show that it is compatible with supersymmetry in the Wess-Zumino gauge.
The conclusion and outlook are shown in section~\ref{sec:summary}.

\section{Yang-Mills gradient flow} 
\label{sec:ymflow}

We review the  gradient flow in four-dimensional Yang-Mills theory, while fixing the  notations used in this paper. 
The gauge group is $SU(N)$ and the group generators $T^a$ ($a=1,\cdots,N^2-1$) are hermitian matrices that 
satisfy the standard relations given in appendix \ref{sec:notation_SUN}. 
The Einstein summation convention is used throughout this paper.

The action of Yang-Mills theory in four dimensional Euclidean space 
whose coordinate is expressed as $x_\mu \, (\mu=0,1,2,3)$ is given by
\begin{eqnarray}
S_{\rm YM}=\frac{1}{2 g^2}\int d^4x\ {\rm tr}\left\{F^2_{\mu\nu}(x)\right\},
\label{def_S}
\end{eqnarray}
where
\begin{eqnarray}
F_{\mu\nu}(x)=\partial_{\mu}A_{\nu}(x)-\partial_{\nu}A_{\mu}(x)+i[A_{\mu}(x),A_{\nu}(x)].
\label{field_tensor}
\end{eqnarray}
The gauge field $A^a_\mu(x)$ is now expressed 
as a matrix-valued field as $A_\mu(x) = \sum_{a=1}^{N^2-1} A_\mu^a(x) T^a$.

To define the gradient flow equation, we introduce a flow-time $t\ (\ge0)$ 
and assume that $A_\mu(x)$ depends on the flow time as $ A_\mu(x) \rightarrow B_\mu(t,x)$ which satisfies
a boundary condition,
\begin{eqnarray}
 B_{\mu}(t,x)\lvert_{t=0}=A_{\mu}(x).\label{BC_Amu}
\end{eqnarray} 
Then the gradient flow equation is formally defined as the gradient of the Yang-Mills action:
\begin{eqnarray}
\partial_t B^a_{\mu}(t,x)= - g^2 \frac{\delta S_{\rm YM}}{\delta A^a_\mu(x)} \bigg\lvert_{A^a_\mu(x) \rightarrow B^a_\mu(t,x)} 
\end{eqnarray}
where $\frac{\delta}{\delta A^a_{\mu}(x)}$ in the right hand side is the functional derivative with respect to $A^a_{\mu}(x)$.
We thus have
\begin{eqnarray}
\partial_t B_{\mu}(t,x) =D_{\nu} G_{\nu\mu}(t,x),
\label{naive_GF_YM}
\end{eqnarray}
where  
\begin{eqnarray}
\begin{split}
&D_\mu \varphi =\partial_\mu  \varphi + i [B_\mu,  \varphi],
\\
& G_{\mu\nu}  = \partial_{\mu} B_{\nu}-\partial_{\nu} B_{\mu}+i[B_{\mu},B_{\nu}].
\end{split}
\label{covariant_derivative_t_dependent _field}
\end{eqnarray}
Throughout this paper, $D_\mu$ that acts on any $t$-dependent field $\varphi(t,x)$ is defined by (\ref{covariant_derivative_t_dependent _field}).

The flow equation is manifestly covariant under a time independent gauge transformation. 
It is known that the gauge covariance becomes an obstacle in the perturbative analyses 
of the flow equation. We introduce a gauge fixing term when performing the perturbative calculations as  
\begin{eqnarray}
\partial_t B_{\mu} = D_{\nu} G_{\nu\mu}+\alpha D_{\mu}\partial_{\nu} B_{\nu}
\label{def_GF_YM}
\end{eqnarray} 
with a positive parameter $\alpha$. 

The solution of the gauge fixed flow equation (\ref{def_GF_YM}) with any $\alpha$ 
is related with the solution with $\alpha=0$ through 
the time dependent gauge transformation:
\begin{eqnarray}
B_{\mu} = \Lambda   \left( B_{\mu}\lvert_{\alpha=0} - i\partial_{\mu}\right)\Lambda^{-1},
\label{time_dep_gauge_a0}
\end{eqnarray}
where 
\begin{eqnarray}
\partial_t \Lambda  = -i\alpha \partial_{\mu} B_{\mu} \Lambda.
\label{time_dep_gauge_a0_cond}
\end{eqnarray}
This fact means that for any gauge invariant operator at $t=0$ 
the two equations (\ref{naive_GF_YM}) and (\ref{def_GF_YM}) are equivalent.      

The flow equation (\ref{def_GF_YM}) itself has a kind of gauge covariance with a time dependent gauge transformation:
\begin{eqnarray}
B_{\mu}^{\omega} = B_{\mu} - D_{\mu}\omega ,
\label{t_dep_sym}
\end{eqnarray}
where $\omega(t,x)$ obeys 
\begin{eqnarray}
\partial_t \omega = \alpha D_{\mu}\partial_{\mu}\omega .
\label{t_dep_sym_2}
\end{eqnarray}
The boundary condition on $\omega$ at $t=0$ can be chosen arbitrarily and the gauge symmetry at $t=0$ remains unfixed.

A surprising result of the Yang-Mills gradient flow is that any correlation function
of the flowed gauge field
is UV finite in all order of perturbation theory once the boundary four-dimensional theory 
is renormalized in the standard way~\cite{Luscher:2011bx}. 
The time dependent gauge symmetry in (\ref{t_dep_sym}) and (\ref{t_dep_sym_2}) then play 
an essential role in the proof of the finiteness of correlation functions. 

\section{Gradient flow equation in ${{\cal N}=1}$ SYM} 
\label{sec:susygf}

We basically follow Ref.~\cite{Wess:1992cp} with the Wick rotation to the Euclidean space. 
The convention of superfields and differential operators after the Wick rotation 
are summarized in appendix \ref{sec:notation_SUSY}.

\subsection{${\cal N}=1$ SYM action and superfield formalism}
\label{subsec:superfield}

The Euclidean action of ${{\cal N}=1}$ SYM  is given by 
\begin{eqnarray}
S=\frac{1}{g^2} \, \int d^4 x \, {\rm tr} \bigg\{
\frac{1}{2} F_{\mu\nu}^2 + 2 i \bar \lambda \bar \sigma_\mu D_\mu \lambda  + D^2
\bigg\}(x)
\label{component_SYM_action}
\end{eqnarray}
where  $\lambda_\alpha (x)$ and $\bar\lambda_{\dot \alpha}(x)$ with $\alpha=1,2$ are two-component spinors,
$A_\mu(x)$ is the gauge field, $D(x)$ is an auxiliary field ($D^a(x) \in \mathbb{R}$). 
Four dimensional sigma matrices are defined as $\sigma_\mu=(-i I,\sigma^i)$ and $\bar\sigma_\mu=(-i I,-\sigma^i)$
with the standard Pauli matrices  $\sigma^i$.
The gauge field tensor $F_{\mu\nu}(x)$ is given by (\ref{field_tensor}) and the covariant derivative $D_\mu$ is defined by
\begin{eqnarray}
D_\mu \varphi(x) =\partial_\mu  \varphi(x) + i [A_\mu(x),  \varphi(x)],
\end{eqnarray} 
for any field $\varphi(x)$ in the adjoint representation of the gauge group.  
The action is invariant under the infinitesimal gauge transformation,
\begin{eqnarray}  
\begin{split}
&
\delta^g_\omega A_\mu(x)  = - D_\mu \omega (x) \\  
& \delta^g_\omega \varphi(x) = i [\omega(x),\varphi(x)].
\end{split}
\label{standard_gauge_transf}
\end{eqnarray}

The supersymmetry transformation is defined by 
\begin{equation}
\begin{split}
&
 \delta_{\xi} A_{\mu} (x) = i\xi{\sigma}_{\mu}\bar{\lambda} (x)
                          +i\bar{\xi}\bar{\sigma}_{\mu}\lambda (x) \\
& \delta_{\xi} \lambda (x) =  \sigma_{\mu\nu}\xi  F_{\mu\nu}(x) -\xi D (x) \\
& \delta_{\xi} \bar\lambda (x) = \bar\sigma_{\mu\nu}\bar\xi  F_{\mu\nu}(x) +\bar\xi D (x) \\
& \delta_{\xi} D (x)       = i\xi\sigma_{\mu}D_{\mu}\bar{\lambda} (x)
                          -i\bar{\xi}\bar{\sigma}_{\mu}D_{\mu}\lambda (x),
\end{split}
\label{DF_super}
\end{equation}
where $\xi_\alpha$ and $\bar\xi_{\dot\alpha}$ are two-component Grassmann parameters that do not depend on $x$. 
We can show that the action is also invariant under this transformation 
using several identities given in appendix \ref{sec:notation_SUSY}.

We now define ${{\cal N}=1}$ SYM in the superfield formalism. 
The supersymmetry transformation of any superfield $ F(x,\theta,\bar\theta) $ is given by
\begin{eqnarray}
\delta^0_\xi F(x,\theta,\bar\theta) = (\xi Q+ \bar \xi \bar Q) \, F(x,\theta, \bar\theta),
\label{linear_super_transf_F}
\end{eqnarray}
where 
\begin{eqnarray}
\begin{split}
&
Q_{\alpha} = 
\frac{\partial}{\partial\theta^{\alpha}} 
-i(\sigma_{\mu})_{\alpha\dot{\alpha}}\bar\theta^{\dot\alpha}\partial_{\mu} \\
& \bar{Q}_{\dot\alpha} = 
-\frac{\partial}{\partial\bar\theta^{\dot\alpha}} 
+i\theta^{\alpha}(\sigma_{\mu})_{\alpha\dot{\alpha}}\partial_{\mu}.
\end{split}
\end{eqnarray}  
The transformation of each component field can be read from (\ref{linear_super_transf_F}). 
Note that 
(\ref{linear_super_transf_F}) is a linear transformation although (\ref{DF_super}) is non-linear 
since  $Q_{\alpha}$ and $\bar{Q}_{\dot\alpha}$ are linear derivative operators. 

The vector superfield $V(x,\theta,\bar{\theta})$ is defined as 
\begin{eqnarray}
&& V(x,\theta,\bar{\theta}) = C(x) + i\theta\eta(x) - i\bar{\theta}\bar{\eta}(x) \nonumber\\
 &&\qquad  +\frac{i}{2}\theta\theta(M(x)+iN(x))
      -\frac{i}{2}\bar\theta\bar\theta(M(x)-iN(x)) \nonumber\\
      &&\qquad -\theta\sigma_{\mu}\bar{\theta}A_{\mu}(x) \nonumber\\
&& \qquad +i\theta\theta\bar{\theta}\left(\bar{\lambda}(x)+\frac{i}{2}\bar{\sigma}_{\mu}\partial_{\mu}\eta(x)\right) \nonumber\\
&& \qquad
      -i\bar{\theta}\bar{\theta}\theta\left(\lambda(x)+\frac{i}{2}\sigma_{\mu}\partial_{\mu}\bar{\eta}(x)\right) \nonumber\\
 && \qquad   +\frac{1}{2}\theta\theta\bar{\theta}\bar{\theta}\left(iD(x)+\frac{1}{2}\Box C(x)\right), 
 \label{vector_superfield}
\end{eqnarray}
where $C, D, M, N$ and $A_{\mu}$ are real fields, and $\eta, \bar\eta, \lambda, \bar\lambda$ are spinor fields.
The superfields $\Phi(x,\theta,\bar{\theta})$ and $\bar \Phi(x,\theta,\bar{\theta})$ that satisfy 
$D_\alpha \bar \Phi =\bar{D}_{\dot\alpha}  \Phi = 0$
are called chiral superfields where
\begin{eqnarray}
\begin{split}
&
 D_{\alpha} =
 \frac{\partial}{\partial\theta^{\alpha}}
+i(\sigma_{\mu})_{\alpha\dot{\alpha}}\bar\theta^{\dot\alpha}\partial_{\mu}\\
& \bar{D}_{\dot\alpha} =
 -\frac{\partial}{\partial\bar\theta^{\dot\alpha}}
-i\theta^{\alpha}(\sigma_{\mu})_{\alpha\dot{\alpha}}\partial_{\mu}
\end{split}
\end{eqnarray}
which are covariant derivatives because they commute with $Q_\beta$ and $\bar Q_{\dot\beta}$.

The  ${{\cal N}=1}$ SYM action is then given by 
\begin{eqnarray}
S_{\rm SYM} &=& -\int d^4 x \ \frac{1}{2g^2} {\rm tr}
    \left(
     \left.W^{\alpha}W_{\alpha}\right\lvert_{\theta\theta}
     +\left.\bar{W}_{\dot\alpha}\bar{W}^{\dot\alpha}\right\lvert_{\bar\theta\bar\theta}     
    \right),\nonumber\\
    \label{superfield_sym_action}
\end{eqnarray}
where
\begin{eqnarray}
\begin{split}
&
W_{\alpha} = -\frac{1}{8}\bar{D}\bar{D} e^{-2V}D_{\alpha} e^{2V} \\
&
 \bar{W}_{\dot\alpha} = \frac{1}{8} D D e^{2V}\bar{D}_{\dot\alpha} e^{-2V}.
\end{split}
\end{eqnarray}
$W_{\alpha}$  and $\bar{W}_{\dot\alpha}$ are chiral superfields 
since $D_\alpha \bar W_{\dot\beta}=\bar D_{\dot\alpha}  W_\beta = 0$. 
The integrals of $\theta\theta$ and $\bar\theta\bar\theta$ components in (\ref{superfield_sym_action}) are 
invariant under linear supersymmetry transformation (\ref{linear_super_transf_F}).

The symmetry of the superfield action (\ref{superfield_sym_action}) is higher than that of (\ref{component_SYM_action}). 
 (\ref{superfield_sym_action}) is indeed invariant under an extended gauge transformation,
\begin{eqnarray}
\begin{split}
& e^{2V^{\prime}} = e^{-i\bar\Lambda}e^{2V} e^{i\Lambda} \\
& e^{-2V^{\prime}} = e^{-i\Lambda}e^{-2V}e^{i\bar\Lambda},
 \end{split}
 \label{egs}
\end{eqnarray}
where $\bar{D}_{\dot\alpha}\Lambda= {D}_{\alpha}\bar\Lambda=0$.  
As the $\theta$ and $\bar\theta$ expansions,
we have  
\begin{eqnarray}
\begin{split}
&
\Lambda(y,\theta)= A(y) + \sqrt{2}\theta\psi(y) + i \theta\theta F(y)
\\
& \bar\Lambda(\bar y, \bar\theta) = \bar A(\bar y) + \sqrt{2} \bar\theta \bar\psi (\bar y) + i \bar\theta \bar\theta \bar F (\bar y)
\end{split}
\end{eqnarray}
where $y_\mu=x_\mu + i \theta \sigma_\mu \bar\theta$ and $\bar y_\mu=x_\mu - i \theta \sigma_\mu \bar\theta$.
We should note that (\ref{egs})  is a non-linear transformation.

We can set 
\begin{equation}
C = \eta = \bar\eta = M = N = 0,
\label{WZ_gauge_fixing}
\end{equation}
choosing the component fields of $\Lambda$ and  $\bar{\Lambda}$ by hand. 
This is so-called Wess-Zumino gauge fixing. 
The vector superfield is then given only by the field variables in (\ref{component_SYM_action}):     
\begin{eqnarray}
V_{\rm WZ}(x,\theta,\bar{\theta}) &=&  -\theta\sigma_{\mu}\bar{\theta}A_{\mu}(x)
    +i\theta\theta\bar{\theta}\bar{\lambda}(x)
      -i\bar{\theta}\bar{\theta}\theta\lambda(x)\nonumber \\
      && +\frac{i}{2}\theta\theta\bar{\theta}\bar{\theta}D(x).
\end{eqnarray}
Consequently, 
\begin{eqnarray}
\begin{split}
W_{\alpha}(y,\theta) &=
 -i\lambda_{\alpha}(y) + i\theta_\alpha D (y) - i (\sigma_{\mu\nu} \theta)_\alpha F_{\mu\nu}(y)\\
& \quad\, +\theta\theta (\sigma_{\mu} D_{\mu}\bar\lambda)_{\alpha}(y)  \\
\bar{W}^{\dot\alpha}(\bar{y},\bar\theta) &= 
i\bar\lambda^{\dot\alpha}(\bar y) + i \bar\theta^{\dot\alpha} D(\bar y) 
+i (\bar\sigma_{\mu\nu} \bar\theta)^{\dot\alpha} F_{\mu\nu}(\bar y)\\
& \quad\, -\bar\theta\bar\theta (\bar\sigma_{\mu} D_{\mu}\lambda)^{\dot\alpha} (\bar y)
\end{split}
\end{eqnarray}
and the superfield action (\ref{superfield_sym_action}) coincides with (\ref{component_SYM_action}).

The original linear supersymmetry transformation $\delta^0_\xi$  breaks the Wess-Zumino gauge. 
Another supersymmetry transformation that keeps the gauge 
is defined by adding an infinitesimal extended gauge transformation $\delta^{\rm gauge}_\Lambda$ to $\delta^0_\xi$ \cite{deWit:1975veh}:
\begin{eqnarray}
\delta_\xi = \delta^0_\xi + \delta^{\rm gauge}_\Lambda.
\label{formal_DF_super}
\end{eqnarray}
The component fields of $\Lambda$ and $\bar{\Lambda}$
are chosen such that $\delta_\xi$ maintains the Wess-Zumino gauge.
Thus we obtain the supersymmetry transformation (\ref{DF_super}) because the non-linear terms of (\ref{DF_super}) 
come from $\delta^{\rm gauge}_\Lambda$.

%
%
%
%
\subsection{Original derivation of the SYM flow equation}
\label{subsec:KO}

We use $t\ (\ge0)$  as a flow time and
define $t$-dependent vector superfield ${\cal V}(t,x,\theta, \bar\theta)$ as
\begin{equation}
{\cal V}(t,x,\theta,\bar\theta) = V(x,\theta,\bar \theta) \bigg\lvert_{\it field \, replacement}
\end{equation}
with {\it the field replacements}, 
\begin{equation}
\begin{split}
& C(x)            \ \rightarrow \  C^\prime (t,x) \\
& \eta(x)        \ \rightarrow \     \eta^\prime(t,x) \\
& \bar\eta(x)  \ \rightarrow \    \bar \eta^\prime(t,x) \\
& M(x)           \ \rightarrow \   M^\prime(t,x) \\
& N(x)           \ \rightarrow \  N^\prime(t,x) \\
& A_\mu(x)     \ \rightarrow \   B_\mu(t,x) \\
& \lambda(x)    \ \rightarrow \   \chi(t,x) \\
& \bar \lambda(x)     \ \rightarrow \ \bar \chi(t,x) \\
& D(x)            \ \rightarrow \ H(t,x)
\end{split}
\end{equation}
and the boundary condition
 ${\cal V}(t,x,\theta,\bar\theta)\lvert_{t=0} = V(x,\theta,\bar\theta)$.

In order to define a gradient flow equation such that
it is covariant under linear supersymmetry and the extended gauge transformations,  
we define the metric that is invariant under these two transformations:
\begin{eqnarray} 
|| \delta V ||^2  = 
 \int d^8 z\ 
 \frac{1}{2}\ 
 {\rm tr}\left(e^{-2V}\delta e^{2V} \cdot 
               e^{-2V}\delta e^{2V}\right),
  \label{norm}            
\end{eqnarray}
where $z=(x,\theta,\bar \theta)$ and $d^8 z \equiv d^4 x d^2\theta d^2\bar{\theta}$.
The metric $g_{ab}$ is read from (\ref{norm}) as $|| \delta V ||^2  =$$  \int d^8 z\ g_{ab}(V) \delta V^a \delta V^b$.  
With $g^{ac}g_{cb}=\delta_{ab}$, we thus have 
\begin{eqnarray}  
g^{ab}(V) = 
  4 \ {\rm tr}\left\{T^a\frac{{\cal L}_V^2}{{\rm cosh}(2 {\cal L}_V)-1}T^b\right\},
\end{eqnarray} 
where ${\cal L}_{A} B \equiv [A,B] $.

The gradient flow equation for ${\cal V}$ is given in terms of superfield as
\begin{eqnarray}
 \partial_t {\cal V}^a(t,z)  
 = \frac{1}{2} g^{ab} \frac{\delta S_{\rm SYM}}{\delta V^b(z)} \bigg\lvert_{V(z) \rightarrow {\cal V}(t,z)}
 \label{formal_GF_SYM}
\end{eqnarray}
where $\delta/\delta V^b(x,\theta,\bar\theta)$ is a functional derivative. 
After a short calculation, (\ref{formal_GF_SYM}) 
 is also expressed as 
\begin{eqnarray}
 \partial_t {\cal V}
 &=& \frac{{\cal L}_V}{1-e^{-2{\cal L}_V}}
 \left(\frac{1}{2}{\cal D}^{\alpha}W{}_{\alpha} \right)
 \nonumber\\&&
   -\frac{{\cal L}_V}{1-e^{2{\cal L}_V}}
    \left(\frac{1}{2} \bar{\cal D}{}_{\dot\alpha}\bar{W}^{\dot\alpha} \right)\bigg\lvert_{V \rightarrow {\cal V}}
    \label{SYM-GFE}
\end{eqnarray}
with 
\begin{eqnarray}
\begin{split}
 {\cal D}^{\alpha}W_{\alpha} 
\aeq D^{\alpha}W_{\alpha}+\{e^{-2V} (D^{\alpha}e^{2V}),  W_{\alpha} \},\\
 \bar{\cal D}{}_{\dot\alpha}\bar{W}^{\dot\alpha} 
\aeq \bar{D}_{\dot\alpha}\bar{W}^{\dot\alpha}+\{e^{2V} (\bar{D}_{\dot\alpha}e^{-2V}), \bar{W}^{\dot\alpha} \},
\end{split}
\end{eqnarray}
which are covariant under super and gauge transformations.

With and without the metric, 
(\ref{formal_GF_SYM})  is covariant under $t$-independent supersymmetry transformation  
which is obtained by replacing $F(z)$ in (\ref{linear_super_transf_F}) with  ${\cal V}(t,z)$
since the both sides are given in terms of superfield. 
However the extended gauge transformation is broken without the metric since it is  non-linear transformation. 
With the metric, 
(\ref{formal_GF_SYM}) is also covariant for the extended gauge transformation 
as in the case of general covariance.

\subsection{The issue of SYM flow equation in the Wess-Zumino gauge}
\label{subsec:KO2}

The Wess-Zumino gauge fixing (\ref{WZ_gauge_fixing}) is broken by the time evolution    
because the breaking term appear in R.H.S of (\ref{formal_GF_SYM}). 
The authors of Ref.\cite{Kikuchi:2014rla} have tried to construct the flow equation in the gauge 
adding a fixing term given by an extended gauge transformation 
to (\ref{formal_GF_SYM}): 
\begin{eqnarray}
\partial_t {\cal V}^a(t,z)  
 = \frac{1}{2} g^{ab} \frac{\delta S_{\rm SYM}}{\delta V^b} + \delta_{\Lambda}^{\rm gauge} V^a \bigg\lvert_{V(z) \rightarrow {\cal V}(t,z)},
 \label{SYM_flow_with_fixing}
\end{eqnarray}
where 
\begin{eqnarray}
\delta_{\Lambda}^{\rm gauge} V =
\frac{{\cal L}_V}{1-e^{-2{\cal L}_V}} \left(
 i\Lambda \right) 
   + \frac{{\cal L}_V}{1-e^{2{\cal L}_V}}
    \left( i \bar\Lambda \right),
\label{ext_gauge}
\end{eqnarray}
which is the infinitesimal form of (\ref{egs}).

In Ref.~\cite{Kikuchi:2014rla}, 
the gauge parameters $\Lambda$ and $\bar\Lambda$ are chosen 
such that the following three conditions hold: \vspace{-2mm} \\

 \hspace{0.5cm}  (i)
 $\Lambda$ and $\bar\Lambda$ are functions of $V$.\\
 \hspace{1cm}  (ii)
 The flow equation is invariant under $\delta_{\xi}^{0}$. \\
 \hspace{1cm}  (iii)
 The WZ gauge is kept at any flow time. \vspace{2mm}  \\
The unique solution of these three conditions is given by
\begin{eqnarray}
\begin{split}
 i\Lambda &= \frac{1}{8}\bar{D}^2(D^2V+2[D^2V,V]) \\
 -i\bar\Lambda &= \frac{1}{8}D^2(\bar{D}^2V-2[\bar{D}^2V,V]).
\end{split}
\label{sol_Lambda}
\end{eqnarray}
Substituting (\ref{sol_Lambda}) into (\ref{SYM_flow_with_fixing}) and setting ${\cal V}$ at  $t=0$ to $V_{\rm WZ}$, 
 the flow equation is defined in terms of component fields.

We make a few comments on the flow equation in the Wess-Zumino gauge 
with (\ref{sol_Lambda}). 
In the case of the Yang-Mills flow (\ref{def_GF_YM}), 
$\alpha$ is a free parameter and the flow equation is gauge covariant if we set $\alpha=0$.
However, such gauge covariance is broken in the present case
due to the choice (\ref{sol_Lambda}). 
With the time dependent gauge transformation (\ref{time_dep_gauge_a0}),  the solution 
with any non-zero $\alpha$ is equivalent to that without the gauge fixing term ($\alpha=0$) 
in the  Yang-Mills flow. 
The corresponding supersymmetric relation is, however, unknown.
Furthermore, it is not clear whether or not the flow equation is consistent with SUSY transformation
in the Wess-Zumino gauge (\ref{DF_super}) \cite{KK_Private}. 

We will reconstruct the flow equation in the Wess-Zumino gauge and study the symmetry properties  in the next section.



\section{New derivation of gradient flow equation in the Wess-Zumino gauge }
\label{subsec:KU}

We propose a method of defining a SYM-gradient flow in the Wess-Zumino gauge  
and show that the constructed flow is supersymmetric in the sense that the
five-dimensional supersymmetry and the flow-time derivative commute 
with each other up to a gauge transformation.

\subsection{Supersymmetric gradient flow}

In the prescription of the Wess-Zumino gauge fixing, 
the linear supersymmetry transformation 
$\delta_{\xi}^{0}$ defined in (\ref{formal_DF_super}) is broken. 
The component fields of chiral superfields, which are parameters of extended gauge transformation, 
are chosen by hand to keep the gauge.
So we can say that  (i) and (ii) in Sec.~\ref{subsec:KO} are too strong conditions which are unnecessary.

We impose only the condition (iii) and choose the component fields of $\Lambda$ and $\bar\Lambda$ by hand
to keep the Wess-Zumino gauge at any flow time: 
\begin{eqnarray}
\begin{split}
  & A =  D -\theta \\
  & \bar A =  -D -\theta \\
  & \psi = \sqrt{2}i\sigma_{\mu}D_{\mu}\bar{\lambda} \\
  & \bar\psi = \sqrt{2}i\bar\sigma_{\mu}D_{\mu}{\lambda} \\
  & F  = \bar F = 0,
  \end{split}
\label{gauge_choice}
\end{eqnarray}
where $\theta^a(t,x)$ is any real function. 
We thus obtain a flow equation in the Wess-Zumino gauge:   
\begin{eqnarray}
\begin{split}
 & \hspace{0mm} 
\partial_t C^\prime =
   \partial_t \eta^\prime
 =\partial_t \bar\eta^\prime   
 = \partial_t {M^\prime} 
 = \partial_t {N^\prime} = 0, 
 \end{split}
 \label{WZ_gauge_keep}
\end{eqnarray}
and
 \begin{eqnarray}
\begin{split}
 &\hspace{0mm} 
\partial_t B_\mu = D_{\nu} G_{\nu\mu} 
             -(\bar{\chi} \bar{\sigma}_{\mu}\chi+\chi\sigma_{\mu}\bar{\chi})
             -D_{\mu}\theta  \\
& \hspace{0mm} 
\partial_t {\chi} = -\sigma_{\mu}\bar{\sigma}_{\nu} D_{\mu}D_{\nu}\chi
               -i[\chi, H] +i[\theta, \chi] \\
& \hspace{0mm} 
\partial_t {\bar{\chi}} = -\bar\sigma_{\mu}\sigma_{\nu}D_{\mu}D_{\nu}\bar{\chi}
                +i[\bar{\chi},H] +i[\theta, \bar{\chi}] \\
& \hspace{0mm} 
\partial_t {H} = D_{\mu}D_{\mu}  H \\& \qquad
 + (D_{\mu}\chi\sigma_{\mu}\bar{\chi}+\bar{\chi}\bar{\sigma}_{\mu}D_{\mu}\chi
     -D_{\mu}\bar{\chi}\bar{\sigma}_{\mu}\chi-\chi\sigma_{\mu}D_{\mu}\bar{\chi})
\\& \qquad +i[\theta,H]. \hspace{-10mm}
 \end{split}
 \label{flow_in_WZ}
\end{eqnarray}
The Wess-Zumino gauge fixing maintains at any flow time as shown in (\ref{WZ_gauge_keep}).
(\ref{gauge_choice}) leads to non-linear couplings of fields in (\ref{flow_in_WZ}).
The flow equation for the fermion is given as $ \slashb{D}^2 \psi$ 
instead of $D_\mu D_\mu \psi$  which is proposed in Ref.~\cite{Luscher:2013cpa}.
The second terms of R.H.S. of $\partial_t B_\mu$ and $\partial_t {H} $
 is actually proportional to the group generator 
noting that $ \bar{\chi}^a \bar{\sigma}_{\mu} \chi^b=-\chi^b \sigma_{\mu} \bar \chi^a$.

We do not have any constraint on $\theta$ that is the real part of $A$ in the Minkowski space, 
which corresponds to the degree of freedom 
of the ordinary gauge transformation (\ref{standard_gauge_transf}).
Thus the Wess-Zumino gauge of SYM-flow equation is independently taken from the ordinary gauge transformation.
Indeed, if we set $\theta=0$, the flow equation is manifestly gauge covariant 
for $t$-independent gauge transformations. 
As in the case of Yang-Mills flow, we can choose any gauge fixing term 
such as $\theta=-\alpha \partial_{\mu}B_{\mu}$ with any  parameter $\alpha$ 
although the flow equation of Ref.~\cite{Kikuchi:2014rla} is given 
with a specific gauge fixing term  $\theta=-\partial_{\mu}B_{\mu}$ ($\alpha=1$).

\subsection{Symmetry of the SYM flow equation}

Let us study the symmetry properties of the flow equation. 
 
As a natural extension, five dimensional supersymmetry can be defined 
replacing $A_{\mu}, \lambda,\bar\lambda, D$ in (\ref{DF_super})
with corresponding flow fields:   
\begin{eqnarray}
\begin{split}
&
 \delta_{\xi} B_{\mu} = i\xi{\sigma}_{\mu}\bar{\chi}
                          +i\bar{\xi}\bar{\sigma}_{\mu}\chi \\
& \delta_{\xi} \chi =  \sigma_{\mu\nu}\xi G_{\mu\nu} - \xi H \\
& \delta_{\xi} \bar\chi = \bar\sigma_{\mu\nu}\bar\xi  G_{\mu\nu} + \bar\xi H \\
& \delta_{\xi} H       = i\xi\sigma_{\mu}D_{\mu}\bar{\chi}
                          -i\bar{\xi}\bar{\sigma}_{\mu} D_{\mu}\chi
\end{split}
\end{eqnarray}
One can naively expect that $\partial_t \delta_\xi= \delta_\xi \partial_t$ 
if supersymmetry and  the flow are consistent with each other.   
 
After some calculations, 
we find that the commutation relation does not hold for $V_{\rm WZ}$ and we actually have
\begin{eqnarray}
 (\partial_t \delta_{\xi} - \delta_{\xi} \partial_t) V_{\rm WZ}  =\delta_{\omega}^{g} V_{\rm WZ}, 
 \label{susy_invariance_g}
\end{eqnarray}
where $\delta_{\omega}^{g}$ represents the ordinary gauge transformation  
with the gauge transformation function, 
\begin{eqnarray}
\omega 
= i D_{\mu}(\xi\sigma_{\mu}\bar\chi+\bar\xi\bar\sigma_{\mu}\chi) 
  +\delta_\xi \theta, 
\end{eqnarray}
where the gauge transformation of flowed fields are  replacing $\omega(x)$ in (\ref{standard_gauge_transf}) by $\omega(t, x)$. 
The relation (\ref{susy_invariance_g}) implies that the flow and supersymmetry 
are consistent with each other for any gauge invariant operator since R.H.S. vanishes. 
In this sense, the constructed flow equation is supersymmetric.

The gauge covariance (\ref{time_dep_gauge_a0}) with (\ref{time_dep_gauge_a0_cond}) also holds 
for the SUSY flow equation (\ref{flow_in_WZ}).
We can show that two solutions of the flow equations with and without $\theta$ are related as
\begin{eqnarray}
\begin{split}
& B_{\mu} = \Lambda   \left( B_{\mu}\lvert_{\theta=0} - i\partial_{\mu}\right)\Lambda^{-1} \\
& \varphi = \Lambda \varphi\lvert_{\theta=0} \Lambda^{-1}, 
\end{split}
\label{SYM_flow_symmetry1}
\end{eqnarray}
where $\partial_t \Lambda  = i \theta \Lambda$ and $\varphi =\chi,\bar\chi, H$.
The existence of this transformation suggests us that we can take any $\theta$
in perturbatve analyses of the flow equation. 
Without fixing the WZ gauge, this kind of relation also holds for \eqref{SYM_flow_with_fixing}:
\begin{eqnarray}
 e^{2\cal V} = e^{-i\bar\Xi}\left.e^{2\cal V}\right\lvert_{\Lambda=0}e^{i\Xi},
 \label{superfield_relation1}
\end{eqnarray} 
where $\Xi$ and $\bar\Xi$ obey 
\begin{eqnarray}
 \partial_t e^{i\Xi} = ie^{i\Xi} \Lambda,\quad  \partial_t e^{-i\bar\Xi} = -i\bar\Lambda e^{-i\bar\Xi}.
 \label{superfield_relation2}
\end{eqnarray}
Note that 
\eqref{SYM_flow_symmetry1}
is also derived from \eqref{superfield_relation1} and \eqref{superfield_relation2} setting the WZ gauge.

We find another relation which is a supersymmetric version of (\ref{t_dep_sym}) and (\ref{t_dep_sym_2}).
The SYM flow equation (\ref{flow_in_WZ}) is covariant under 
a time dependent gauge transformation, 
\begin{eqnarray}
\begin{split}
B_\mu^{\omega} = B_{\mu} - D_{\mu}\omega, \\
\varphi^{\omega} =\varphi + i[\omega, \varphi], 
\end{split}
\label{SYM_flow_symmetry2}
\end{eqnarray}
where $\omega(t,x)$ obeys 
$\partial_t \omega = \delta^g_{\omega} \theta + i[\theta, \omega]$.
Here $\theta$ is assumed to be a function of the component fields, such as 
$\theta=-\alpha \partial_\mu B_\mu$. 
For the superfields, it is easily shown that ${\cal V}_{\Xi}$ defined by
\begin{eqnarray}
 e^{2{\cal V}_{\Xi}} = e^{2{\cal V}} -i\bar \Xi e^{2{\cal V}} +i e^{2{\cal V}} \Xi,
 \label{superfield_relation3}
\end{eqnarray}
is another solution of the fixed equation \eqref{SYM_flow_with_fixing} for 
infinitesimal transformation parameters $\Xi,\bar \Xi$ satisfying 
\begin{eqnarray}
\begin{split}
\partial_t \Xi = i [\Xi, \Lambda] + \delta_\Xi \Lambda, \\
\partial_t \bar \Xi = i [\bar \Xi, \bar \Lambda] + \delta_\Xi \bar \Lambda,
\end{split}
\label{superfield_relation4}
\end{eqnarray}
where $\delta_\Xi \Lambda = \Lambda(V_\Xi) - \Lambda(V)$.  
Note again that 
\eqref{SYM_flow_symmetry2}
is derived from \eqref{superfield_relation3} and \eqref{superfield_relation4} setting the WZ gauge.

\section{Summary}
\label{sec:summary}

We have defined a gradient flow equation in ${\cal N}=1$ supersymmetric 
Yang-Mills theory in terms of the component fields of the Wess-Zumino gauge.
We have shown that the obtained flow is consistent with
supersymmetry (de Wit-Freedman transformation) such that 
the commutator of time-derivative and  five-dimensional  SUSY transformation vanishes 
up to a gauge transformation. In this sense, one can say that the flowed gauge multiplet in 
the Wess-Zumino gauge remains as a supermultiplet at non-zero flow time.

Although the flowed adjoint fermion and the auxiliary field receive extra renormalizations 
for the non-SUSY flow,  
one can expect that there are no such extra renormalizations in the supersymmetric flow.
This is the direct consequence that the renormalization property of the gauge field 
 is extended to ${\cal N}=1$ gauge multiplet thanks to five-dimensional supersymmetry. 
One can also expect that the energy-momentum tensor, supercurrent and R-current 
form a five-dimensional Ferrara-Zumino multiplet.
The SUSY flow is thus very attractable, but
the actual renormalization property is still not clear and further studies are needed.

The SUSY gradient flow of the same kind can be defined in various SUSY models.  
The method presented in this paper could be very useful to build the theory of SUSY flow
in ${\cal N}=1$ SQCD and ${\cal N}=2$ theories. 
But it would not be straightforward for ${\cal N}=4$ SYM 
since ${\cal N}=4$ theory has no off-shell formulation which is important 
in showing that the flow is supersymmetric.
The SUSY flow can also be given for the low-dimensional SUSY theories and the Wess-Zumino model.
We can say that the SUSY flow is a systematic and promising approach 
and will be widely used in the actual lattice computations in the near future.

\acknowledgement
{\it Acknowledgements}:
We would like to thank Masafumi Fukuma, Nobuhito Maru and Hiroshi Suzuki for their encouraging comments.
D.K. would also like to thank Kengo Kikuchi for helpful private discussions on supersymmetric gradient flow. 
This work is supported by JSPS KAKENHI Grant Numbers JP16K05328, JP16K13798, JP19K03853,	JP20K03924
and JP22H01222.

\begin{appendix}

\section{Generators of $SU(N)$ gauge group}
\label{sec:notation_SUN}

The Lie algebra of $SU(N)$ group is the $N^2-1$ dimensional linear space spanned by the generators $T^a (a=1,2, \cdots,N^2-1)$,
\begin{eqnarray}
 (T^a)^{\dagger}=T^a
\end{eqnarray}
with the standard commutation relation,
\begin{eqnarray}
[T^a,T^b]=if^{abc}T^c,
\end{eqnarray}
where the structure constant $f^{abc}$ is real and totally anti-symmetric. 

In the fundamental representation, $T^a$ satisfy the normalization
\begin{eqnarray}
 {\rm tr}(T^aT^b)=\frac{1}{2}\delta^{ab}
\end{eqnarray}
and we have a useful relation, 
\begin{eqnarray}
\sum_{a=1}^{N^2-1}{\rm tr}(AT^a) {\rm tr}(BT^a) = \frac{1}{2}{\rm tr}(AB),
\end{eqnarray} 
for any traceless matrices $A$ and $B$.


\section{Supersymmetry in four dimensional Euclidean space}
\label{sec:notation_SUSY}

We basically follow the convention of Ref.~\cite{Wess:1992cp} 
with the Wick rotation to four-dimensional Euclidean space. 
The Minkowski time $t$ is replaced as $t \rightarrow -it$  
where $t=x^0$ and $\partial_t \rightarrow i \partial_t$.
The gauge field $A_0$ and the auxiliary fields $X$ are then replaced as
$A_0 \rightarrow iA_0, X \rightarrow i X$.  
After these replacements we identify the Minkowski action 
$S_{(M)}$ as the Euclidean action $iS_{(E)}$.

\subsection{Convention and spinor algebra in Euclidean space}
\label{sec:notation_alg}

In the Euclidean space $\psi_\alpha$ and $\bar\psi_{\dot\alpha}$ ($\alpha=1,2$) 
transform as independent spinors under $SU(2)_R$ and $SU(2)_L$ group 
and they are not related with each other under the complex conjugate.
We define  the invariant tensors of  $SU(2)_R$ and $SU(2)_L$ as 
\begin{eqnarray}
 \epsilon_{21}=\epsilon^{12}=\epsilon_{\dot2\dot1}=\epsilon^{\dot1\dot2}=1,\quad
  \epsilon_{12}=\epsilon^{21}=\epsilon_{\dot1\dot2}=\epsilon^{\dot2\dot1}=-1
\nonumber\\
\end{eqnarray}
with the others are zero. 
Note that
$\epsilon_{\alpha\beta}\epsilon^{\beta\gamma}=\delta_{\alpha}{}^{\gamma}$ and  
$\epsilon_{\dot\alpha\dot\beta}\epsilon^{\dot\beta\dot\gamma}=\delta_{\dot\alpha}{}^{\dot\gamma}$.
Then
\begin{eqnarray}
\psi \chi \equiv \psi^\alpha \chi_\alpha, 
\qquad  \bar\psi \bar\chi \equiv  \bar\psi_{\dot\alpha} \bar\chi^{\dot\alpha}
\label{contraction}
\end{eqnarray}
are Lorenz scalars.  
We use a totally anti-symmetric tensor $\epsilon_{\mu\nu\rho\sigma}$ with $\epsilon_{0123}=-1$.

The four dimensional sigma matrices in the Euclidean space $(\sigma_{\mu})_{\alpha\dot{\beta}}$ and $(\bar\sigma_{\mu})^{\dot\alpha \beta}$ 
are defined as
\begin{eqnarray}
&& \sigma_0 =
\left(
\begin{array}{cc} -i &  0    \\  0 &  -i   \end{array}
\right),
\qquad  
 \sigma_1 =
\left(
\begin{array}{cc}  0 &  1    \\  1 &  0   \end{array}
\right)  \nonumber \\ 
&& \sigma_2 =
\left(
\begin{array}{cc}  0 &  -i    \\  i &  0   \end{array}
\right),
\qquad  
 \sigma_3 =
\left(
\begin{array}{cc}  1 &  0    \\  0 &  -1   \end{array}
\right)
 \\
&& (\bar{\sigma}_{\mu})^{\dot{\alpha}\alpha} = \epsilon^{\dot\alpha\dot\beta}\epsilon^{\alpha\beta}
 (\sigma_{\mu})_{\beta\dot{\beta}}
\nonumber \\
&& \bar \sigma_0 = \sigma_0, \qquad \bar\sigma_i=-\sigma_i \ \ (i=1,2,3). \nonumber  
\end{eqnarray}
Note that the matrix elements of $\sigma_\mu$ are different from those of Minkowski's $\sigma_\mu$. 
\footnote{They are related as
$\sigma^{(E)}_0 =-i\sigma_0^{(M)}$, $\bar\sigma^{(E)}_0 =-i\bar\sigma_0^{(M)}$, and 
$\sigma^{(E)}_j =\sigma_j^{(M)}$, $\bar\sigma^{(E)}_j =\bar\sigma_j^{(M)}$ for $j=1,2,3$.
}
As with (\ref{contraction}),  the contractions of
$\chi \sigma_\mu \bar\psi$  and 
$\bar\chi \bar\sigma_\mu \psi$  mean  
$\chi^\alpha (\sigma_\mu)_{\alpha\dot\beta} \bar\psi^{\dot\beta}$ 
and 
$\bar\chi_{\dot\alpha} (\bar\sigma_\mu)^{\dot\alpha\beta} \psi_{\beta}$, respectively.

We present the useful identities for sigma matrices,
\begin{equation}
\begin{split}
&{\rm Tr}(\sigma_\mu \bar\sigma_\nu) = -2  \delta_{\mu\nu}  \\
& (\sigma_{\mu})_{\alpha\dot\alpha}(\bar\sigma_{\mu})^{\dot\beta\beta}  =-2\delta_{\alpha}{}^{\beta}\delta_{\dot\alpha}{}^{\dot\beta},
\end{split}
\end{equation}
\begin{equation}
\begin{split}
 (\sigma_{\mu}\bar{\sigma_{\nu}}+\sigma_{\nu}\bar\sigma_{\mu})_{\alpha}{}^{\beta} &=
 -2\delta_{\mu\nu}\delta_{\alpha}{}^{\beta} \\
 (\bar\sigma_{\mu}\sigma_{\nu}+\bar\sigma_{\nu}\sigma_{\mu})^{\dot\alpha}{}_{\dot\beta} &=
 -2\delta_{\mu\nu}\delta^{\dot\alpha}{}_{\dot\beta},
\end{split}
\end{equation}
%
%
and the generators of $SO(4)$ in the spinor representation,
\begin{equation}
\begin{split}
& (\sigma_{\mu\nu})_{\alpha}{}^{\beta} =
  \frac{1}{4}(\sigma_{\mu}\bar{\sigma_{\nu}}-\sigma_{\nu}\bar\sigma_{\mu})_{\alpha}{}^{\beta}  \\
 & (\bar\sigma_{\mu\nu})^{\dot\alpha}{}_{\dot\beta} =
  \frac{1}{4}(\bar\sigma_{\mu}\sigma_{\nu}-\bar\sigma_{\nu}\sigma_{\mu})^{\dot\alpha}{}_{\dot\beta}.
  \end{split}
\end{equation}
The identities for $\sigma_{\mu\nu}$ and two and three $\sigma$ are 
%
%
\begin{equation}
\begin{split}
& (\sigma_{\mu\nu})_{\alpha}{}^{\alpha} = (\bar\sigma_{\mu\nu})^{\dot\alpha}{}_{\dot\alpha} = 0  \\
& (\sigma_{\mu\nu})_{\alpha}{}^\gamma \epsilon_{\gamma \beta} =  (\sigma_{\mu\nu})_{\beta}{}^\gamma \epsilon_{\gamma \alpha} \\
& (\bar\sigma_{\mu\nu})^{\dot\alpha}{}_{\dot\gamma} \epsilon^{\dot\gamma \dot\beta} 
=  (\bar\sigma_{\mu\nu})^{\dot\beta}{}_{\dot\gamma} \epsilon^{\dot\gamma \dot\alpha}.\\
  \end{split}
\end{equation}

%
%
\begin{equation}
\begin{split}
& \epsilon_{\mu\nu\rho\sigma} \, \sigma_{\rho\sigma} = 2\sigma_{\mu\nu} \\
 & \epsilon_{\mu\nu\rho\sigma} \bar\sigma_{\rho\sigma} = -2 \bar\sigma_{\mu\nu}.
  \end{split}
\end{equation}

\begin{equation}
\begin{split}
& \left( \sigma_{\mu} \bar\sigma_{\nu} \right)_{\alpha}{}^{\beta}
  = 2\left( \sigma_{\mu\nu} \right)_{\alpha}{}^{\beta} -\delta_{\mu\nu}\delta_{\alpha}{}^{\beta}\\
& \left( \bar\sigma_{\mu} \sigma_{\nu} \right)^{\dot\alpha}{}_{\dot\beta}
  = 2\left( \bar\sigma_{\mu\nu} \right)^{\dot\alpha}{}_{\dot\beta} -\delta_{\mu\nu}\delta^{\dot\alpha}{}_{\dot\beta}.
\end{split}
\end{equation}

\begin{equation}
\begin{split}
& {\rm Tr}\left( \sigma_{\mu\nu} \sigma_{\rho\sigma} \right)
  = -\frac{1}{2}\delta_{\mu\rho}\delta_{\nu\sigma} +\frac{1}{2}\delta_{\mu\sigma}\delta_{\nu\rho}
    -\frac{1}{2}\epsilon_{\mu\nu\rho\sigma} \\
& {\rm Tr}\left( \bar\sigma_{\mu\nu} \bar\sigma_{\rho\sigma} \right)
  = -\frac{1}{2}  \delta_{\mu\rho}\delta_{\nu\sigma} + \frac{1}{2}\delta_{\mu\sigma}\delta_{\nu\rho}
    +\frac{1}{2}\epsilon_{\mu\nu\rho\sigma}.
\end{split}
\end{equation}

\begin{equation}
\begin{split}
& \sigma_a \bar\sigma_b \sigma_c + \sigma_c \bar\sigma_b \sigma_a
  = 2\left( \delta_{ac}\sigma_b -\delta_{bc}\sigma_a -\delta_{ab}\sigma_c \right) \\
& \bar\sigma_a \sigma_b \bar\sigma_c + \bar\sigma_c \sigma_b \bar\sigma_a
  = 2\left( \delta_{ac}\bar\sigma_b -\delta_{bc}\bar\sigma_a -\delta_{ab}\bar\sigma_c \right).
\end{split}
\end{equation}

\begin{equation}
\begin{split}
& \sigma_a \bar\sigma_b \sigma_c - \sigma_c \bar\sigma_b \sigma_a
  = 2\epsilon_{abcd} \sigma_d \\
& \bar\sigma_a \sigma_b \bar\sigma_c - \bar\sigma_c \sigma_b \bar\sigma_a
  = -2\epsilon_{abcd} \bar\sigma_d.
\end{split}
\end{equation}

%
%
%
\begin{equation}
\begin{split}
& (\sigma_{\mu})_{\alpha\dot\alpha}  (\sigma_{\nu})_{\beta\dot\beta} 
-(\sigma_{\nu})_{\alpha\dot\alpha}  (\sigma_{\mu})_{\beta\dot\beta}\\&\qquad
 = 2 (\sigma_{\mu\nu} \epsilon)_{\alpha\beta} \epsilon_{\dot \alpha \dot\beta} 
+2(\epsilon\bar\sigma_{\mu\nu})_{\dot\alpha\dot\beta} \epsilon_{\alpha\beta} \\
& (\sigma_{\mu})_{\alpha\dot\alpha}  (\sigma_{\nu})_{\beta\dot\beta} 
+(\sigma_{\nu})_{\alpha\dot\alpha}  (\sigma_{\mu})_{\beta\dot\beta}\\&\qquad
 = -\delta_{\mu\nu} \epsilon_{\alpha \beta} \epsilon_{\dot \alpha \dot\beta}
 +4 (\epsilon\bar\sigma_{\rho\mu})_{\dot\alpha\dot\beta}  (\sigma_{\rho\nu} \epsilon)_{\alpha\beta}. 
%
%
  \end{split}
\end{equation}

The results of spinor algebra are given as follows:
\begin{equation}
\begin{split}
& \theta^\alpha \theta^\beta = -\frac{1}{2} \epsilon^{\alpha\beta} \theta \theta\\
& \theta_\alpha \theta_\beta = \frac{1}{2} \epsilon_{\alpha\beta} \theta \theta\\
& \bar\theta^{\dot\alpha} \bar\theta^{\dot\beta} = \frac{1}{2} \epsilon^{\dot\alpha\dot\beta} \bar\theta \bar\theta\\
& \theta_{\dot\alpha} \theta_{\dot\beta} = -\frac{1}{2} \epsilon_{\dot\alpha\dot\beta} \bar\theta \bar\theta.
  \end{split}
\end{equation}

\begin{equation}
\begin{split}
& (\theta \sigma_\mu \bar\theta) (\theta \sigma_\nu \bar\theta) 
= -\frac{1}{2} \theta \theta \bar\theta \bar\theta \delta_{\mu\nu}.
\end{split}
\end{equation}

\begin{equation}
\begin{split}
& \epsilon^{\alpha\beta} \frac{\partial} {\partial \theta^\beta} 
= - \frac{\partial} {\partial \theta_\alpha} \\ 
& \epsilon_{\dot\alpha\dot\beta} \frac{\partial} {\partial \bar\theta_{\dot\beta}} 
= - \frac{\partial} {\partial \bar\theta^{\dot\alpha}}. 
  \end{split}
\end{equation}

\begin{equation}
\begin{split}
& \epsilon^{\alpha\beta} \frac{\partial} {\partial \theta^\alpha}  \frac{\partial} {\partial \theta^\beta} \theta\theta
= 4 \\ 
& \epsilon_{\dot\alpha\dot\beta} \frac{\partial} {\partial \bar\theta_{\dot\alpha}} \frac{\partial} {\partial \bar\theta_{\dot\beta}} 
\bar\theta \bar\theta
= 4. 
  \end{split}
\end{equation}

For anti-commuting spinors $\phi, \psi, \bar\phi, \bar\psi$, we have
\begin{equation}
\begin{split}
& (\theta \phi)( \theta \psi) = -\frac{1}{2} (\theta \theta) (\phi \psi) \\
& (\bar\theta \bar\phi)(\bar\theta \bar\psi)
 = -\frac{1}{2} (\bar\theta \bar\theta) (\bar\phi \bar\psi).
  \end{split}
\end{equation}

\begin{equation}
\begin{split}
& (\theta \phi)(\bar\theta \bar\psi)
 = \frac{1}{2} (\theta \sigma_{\mu}\bar\theta) (\bar\psi \bar\sigma_{\mu} \phi).
  \end{split}
\end{equation}

\begin{equation}
\begin{split}
& (\phi \sigma_{\mu} \bar\psi) = -(\bar\psi \bar\sigma_{\mu} \phi).
  \end{split}
\end{equation}

\begin{equation}
\begin{split}
& (\phi \sigma_{\mu} \bar\sigma_{\nu} \psi) = (\psi \sigma_{\nu} \bar\sigma_{\mu} \phi) \\
& (\bar\phi \bar\sigma_{\mu} \sigma_{\nu} \bar\psi) = (\bar\psi \bar\sigma_{\nu} \sigma_{\mu} \bar\phi).
  \end{split}
\end{equation}

\begin{equation}
\begin{split}
& (\psi\sigma_{\mu} \bar\theta) (\bar\theta \bar\phi) = -\frac{1}{2}\bar\theta \bar\theta (\psi\sigma_{\mu} \bar\phi) \\
& (\theta \sigma_{\mu} \bar\psi)  (\theta \phi) = -\frac{1}{2}\theta \theta (\phi \sigma_{\mu} \bar\psi).
\end{split}
\end{equation}

\begin{equation}
\begin{split}
& (\theta \sigma_\nu \bar\theta) (\psi \sigma_\mu \bar\theta)
= \frac{1}{2} \bar\theta \bar\theta (\theta \sigma_{\nu} \bar\sigma_{\mu} \psi) \\
& (\theta \sigma_\nu \bar\theta) (\theta \sigma_\mu \bar\psi) 
= \frac{1}{2} \theta \theta (\bar\theta \bar\sigma_{\nu} \sigma_{\mu}\bar\psi).
\end{split}
\end{equation}

For matrix-valued spinor fields such as $\psi= \psi^a T^a$ and $\lambda= \lambda^a T^a$,  
the following identities are also useful:

\begin{equation}
\begin{split}
& [ \theta \lambda, \theta \psi ] = \frac{1}{2}\theta\theta (\psi\lambda -\lambda\psi) \\
& [ \bar\theta \bar\lambda, \bar\theta \bar\psi ] = \frac{1}{2}\bar\theta \bar\theta (\bar\psi \bar\lambda -\bar\lambda \bar\psi).
\end{split}
\end{equation}

\begin{equation}
\begin{split}
& [ \bar\theta \bar\lambda, \theta \psi ] = \frac{1}{2}(\theta \sigma_{\mu} \bar\theta) (\psi \sigma_{\mu}\bar\lambda +\bar\lambda \bar\sigma_{\mu} \psi).
\end{split}
\end{equation}

\begin{equation}
\begin{split}
& [ \theta \lambda, \theta \sigma_{\mu} \bar\psi ] = -\frac{1}{2}\theta\theta (\lambda \sigma_{\mu} \bar\psi + \bar\psi \bar\sigma_{\mu} \lambda) \\
& [ \bar\theta \bar\lambda, \bar\theta \bar\sigma_{\mu} \psi ] = -\frac{1}{2}\bar\theta \bar\theta (\bar\lambda \bar\sigma_{\mu} \psi + \psi \sigma_{\mu} \bar\lambda).
\end{split}
\end{equation}

Hereafter, for two-component Grassmann parameters $\eta_\alpha$ and $\xi_\alpha$,
\begin{equation}
\begin{split}
& [ \eta \lambda, \xi \psi ] +[ \xi \lambda, \eta \psi ] 
= (\eta \xi) (\psi \lambda - \lambda \psi) \\
& [ \bar\eta \bar\lambda, \bar\xi \bar\psi ] 
+[ \bar\xi \bar\lambda, \bar\eta \bar\psi ] 
= (\bar\eta \bar\xi) (\bar\psi \bar\lambda - \bar\lambda \bar\psi).
\end{split}
\end{equation}

\begin{equation}
\begin{split}
& (\sigma_{\mu\nu} \xi)_\alpha
(\lambda \sigma_\nu \bar\lambda + \bar\lambda \bar\sigma_\nu \lambda)
=\frac{1}{2}[\xi \lambda, (\sigma_\mu \bar\lambda)_\alpha]
+\frac{1}{2}[\xi \sigma_\mu \bar\lambda, \lambda_\alpha] \\
& (\bar\sigma_{\mu\nu} \bar\xi)^{\dot\alpha}
(\lambda \sigma_\nu \bar\lambda + \bar\lambda \bar\sigma_\nu \lambda)
=\frac{1}{2}[\bar\xi \bar\lambda, (\bar\sigma_\mu \lambda)^{\dot\alpha}]
+\frac{1}{2}[\bar\xi \bar\sigma_\mu \lambda, \bar\lambda^{\dot\alpha}].
\end{split}
\end{equation}

\begin{equation}
\begin{split}
& [\xi \sigma_\nu \bar\psi, (\sigma_{\mu\nu} \lambda)_{\alpha}]
=\frac{1}{2}[(\sigma_\mu \bar\psi)_{\alpha}, \xi \lambda]
+\frac{1}{2}\xi_{\alpha} (\bar\psi \bar\sigma_\mu \lambda
+ \lambda \sigma_\mu \bar\psi) \\
& [\bar\xi \bar\sigma_\nu \psi, (\sigma_{\mu\nu} \lambda)_{\alpha}]
=\frac{1}{2}(\sigma_\mu \bar\xi)_{\alpha} (\lambda \psi - \psi \lambda)
+\frac{1}{2}[\bar\xi \bar\sigma_\mu \lambda, \psi_\alpha].
\end{split}
\end{equation}

\begin{equation}
\begin{split}
& [\xi \sigma_\nu \bar\psi, (\bar\sigma_{\mu\nu}\bar\lambda)^{\dot\alpha}]
=\frac{1}{2}(\bar\sigma_\mu \xi)^{\dot\alpha} (\bar\lambda \bar\psi - \bar\psi \bar\lambda)
+\frac{1}{2}[\xi \sigma_\mu \bar\lambda, \bar\psi^{\dot\alpha}]\\
& [\bar\xi \bar\sigma_\nu \psi, (\bar\sigma_{\mu\nu} \bar\lambda)^{\dot\alpha}]
=\frac{1}{2}[(\bar\sigma_\mu \psi)^{\dot\alpha}, \bar\xi \bar\lambda]
+\frac{1}{2}\bar\xi^{\dot\alpha} (\psi \sigma_\mu \bar\lambda
+ \bar\lambda \bar\sigma_\mu \psi).
\end{split}
\end{equation}

%
%
%
%
\subsection{Chiral and vector superfields}

The superfield $F$ is defined as a Lorentz covariant function of $x,\theta,\bar\theta$ with supersymmetry transformation,
\begin{eqnarray}
\delta_\xi F(x,\theta,\bar\theta) = (\xi Q+ \bar \xi \bar Q) \, F(x,\theta, \bar\theta),
\label{linear_super}
\end{eqnarray}
where $\xi_\alpha$ and $\bar\xi_{\dot\alpha}$ are two component Grassmann global parameters 
and the difference operators $Q_\alpha, \bar Q_{\dot\alpha}$ are defined by
\begin{eqnarray}
\begin{split}
& Q_{\alpha} = 
\frac{\partial}{\partial\theta^{\alpha}} 
-i(\sigma_{\mu})_{\alpha\dot{\alpha}}\bar\theta^{\dot\alpha}\partial_{\mu} \\
& \bar{Q}_{\dot\alpha} = 
-\frac{\partial}{\partial\bar\theta^{\dot\alpha}} 
+i\theta^{\alpha}(\sigma_{\mu})_{\alpha\dot{\alpha}}\partial_{\mu}.
\end{split}
\end{eqnarray} 
The associated super covariant derivatives  which commute with $Q_{\alpha}$ and $\bar{Q}_{\dot\alpha}$ are defined as
\begin{eqnarray}
\begin{split}
& D_{\alpha} =
 \frac{\partial}{\partial\theta^{\alpha}}
+i(\sigma_{\mu})_{\alpha\dot{\alpha}}\bar\theta^{\dot\alpha}\partial_{\mu}  \\
& \bar{D}_{\dot\alpha} =
 -\frac{\partial}{\partial\bar\theta^{\dot\alpha}}
-i\theta^{\alpha}(\sigma_{\mu})_{\alpha\dot{\alpha}}\partial_{\mu}.
\end{split}
\end{eqnarray}
These difference operators obey 
\begin{eqnarray}
\begin{split}
& \{Q_{\alpha},\bar{Q}_{\dot{\alpha}}\} = 2i(\sigma_{\mu})_{\alpha\dot{\alpha}}\partial_{\mu}  \\
& \{D_{\alpha},\bar{D}_{\dot{\alpha}}\} = -2i(\sigma_{\mu})_{\alpha\dot{\alpha}}\partial_{\mu}
\end{split}
\end{eqnarray}
and the other anti-commutation relations vanish.

The supersymmetry transformation of each component field 
is obtained by expanding $F(x,\theta,\bar\theta) $ with respect to $\theta$ and $\bar\theta$  
and comparing the coefficients of the $\theta$-expansion between the both sides of  (\ref{linear_super}).

The chiral and anti-chiral superfields $\Phi(x,\theta,\bar\theta)$ and $\bar\Phi(x,\theta,\bar\theta)$ are defined as superfields that satisfy $\bar D_{\dot\alpha}\Phi = 0$ and $D_{\alpha} \bar\Phi =0$, 
respectively. They are expanded in $\theta$ and $\bar\theta$ as
\begin{eqnarray}
\Phi(x,\theta,\bar\theta) \aeq
 A(x) +i\theta\sigma_{\mu}\bar\theta \partial_{\mu} A(x) +\frac{1}{4}\theta \theta \bar\theta \bar\theta \Box A(x) \nonumber\\
 && +\sqrt{2}\theta\psi(x) -\frac{i}{\sqrt{2}}\theta\theta \partial_{\mu}\psi(x)\sigma_{\mu}\bar\theta 
                          +i\theta\theta F(x),\nonumber\\ \\
\bar\Phi(x,\theta,\bar\theta) \aeq
 \bar A(x) -i\theta\sigma_{\mu}\bar\theta \partial_{\mu} \bar A(x) +\frac{1}{4}\theta \theta \bar\theta \bar\theta \Box \bar A(x) \nonumber\\
 && +\sqrt{2}\bar\theta \bar\psi(x) +\frac{i}{\sqrt{2}}\bar\theta\bar\theta \theta \sigma_{\mu}\partial_{\mu}\bar\psi(x) 
                          +i\bar\theta\bar\theta \bar F(x),\nonumber \\
\end{eqnarray}
where $A, \bar A, F$ and $\bar F$ are independent complex bosonic fields, and $\psi, \bar\psi$ are two component fermions. 
Although 
$\bar\Phi$ is complex conjugate of $\Phi$ in the Minkowski space, that relation is broken 
in the Euclidean space.

Introducing new coordinate $(y, \theta, \bar\theta)$ with $y_{\mu}=x_{\mu}+i\theta\sigma_{\mu}\bar\theta$, 
the derivative operators and $\Phi$ are expressed as 
\begin{eqnarray}
\begin{split}
& Q_{\alpha} = 
\frac{\partial}{\partial\theta^{\alpha}} \\
& \bar{Q}_{\dot\alpha} = 
-\frac{\partial}{\partial\bar\theta^{\dot\alpha}} 
+2i\theta^{\alpha}(\sigma_{\mu})_{\alpha\dot{\alpha}}\partial_{\mu}  \\
& D_{\alpha} =
 \frac{\partial}{\partial\theta^{\alpha}}
+2i(\sigma_{\mu})_{\alpha\dot{\alpha}}\bar\theta^{\dot\alpha}\partial_{\mu} \\
& \bar{D}_{\dot\alpha} =
 -\frac{\partial}{\partial\bar\theta^{\dot\alpha}} \\
& \Phi(y,\theta) = A(y)+\sqrt{2}\theta\psi(y)+i\theta\theta F(y),
\end{split}
\end{eqnarray}
while in $(\bar{y}, \theta, \bar\theta)$ with $\bar{y}_{\mu}=x_{\mu}-i\theta\sigma_{\mu}\bar\theta$,  
\begin{eqnarray}
\begin{split}
& Q_{\alpha} = 
\frac{\partial}{\partial\theta^{\alpha}} 
-2i(\sigma_{\mu})_{\alpha\dot{\alpha}}\bar\theta^{\dot\alpha}\partial_{\mu}  \\
& \bar{Q}_{\dot\alpha} = 
-\frac{\partial}{\partial\bar\theta^{\dot\alpha}}  \\
& D_{\alpha} =
 \frac{\partial}{\partial\theta^{\alpha}}  \\
& \bar{D}_{\dot\alpha} =
 -\frac{\partial}{\partial\bar\theta^{\dot\alpha}}  
 -2i\theta^{\alpha}(\sigma_{\mu})_{\alpha\dot{\alpha}}\partial_{\mu}  \\
& \bar\Phi(\bar{y},\bar\theta) = \bar A(\bar{y})+\sqrt{2}\bar\theta\bar\psi(\bar{y})+i\bar\theta\bar\theta \bar F(\bar{y}).
\end{split}
\end{eqnarray}
Note that $\bar y$ is not a complex conjugate of $y$ in the Euclidean space.

The vector superfield is defined as
\begin{eqnarray}
&& V(x,\theta,\bar{\theta}) = C(x) + i\theta\eta(x) - i\bar{\theta}\bar{\eta}(x) \nonumber\\
 &&\qquad  +\frac{i}{2}\theta\theta(M(x)+iN(x))
      -\frac{i}{2}\bar\theta\bar\theta(M(x)-iN(x))\nonumber\\ 
 &&\qquad-\theta\sigma_{\mu}\bar{\theta}A_{\mu}(x) \nonumber\\
&& \qquad +i\theta\theta\bar{\theta}\left(\bar{\lambda}(x)+\frac{i}{2}\bar{\sigma}_{\mu}\partial_{\mu}\eta(x)\right)\nonumber\\
&&\qquad
      -i\bar{\theta}\bar{\theta}\theta\left(\lambda(x)+\frac{i}{2}\sigma_{\mu}\partial_{\mu}\bar{\eta}(x)\right) \nonumber\\
 && \qquad   +\frac{1}{2}\theta\theta\bar{\theta}\bar{\theta}\left(iD(x)+\frac{1}{2}\Box C(x)\right), 
 \label{vector_superfield}
\end{eqnarray}
where $C, D, M, N$ and $A_{\mu}$ are real bosonic fields, and $\eta, \bar\eta, \lambda, \bar\lambda$ are spinor fields.
The Euclidean vector superfield (\ref{vector_superfield}) is obtained by the Wick rotation of 
vector superfield  in the Minkowski space $V_M$ which satisfies $V_M^\dag=V_M$.

\subsection{${\cal N}=1$ super Yang-Mills in four dimensional Euclidean space}
\label{sec:notation_SYM}


In the gauge theory, the vector superfield is a matrix-valued field as  
\begin{eqnarray}
&& V = V^a T^a.
\end{eqnarray}
The $N=1$ SYM action is given in terms of vector superfield:
\begin{eqnarray}
S_{\rm SYM} \aeq -\int d^4 x \ \frac{1}{2g^2} {\rm tr}
    \left(
     \left.W^{\alpha}W_{\alpha}\right\lvert_{\theta\theta}
     +\left.\bar{W}_{\dot\alpha}\bar{W}^{\dot\alpha}\right\lvert_{\bar\theta\bar\theta}     
    \right),\nonumber \\
\end{eqnarray}
where
\begin{eqnarray}
\begin{split}
& W_{\alpha} = -\frac{1}{8}\bar{D}\bar{D} e^{-2V}D_{\alpha} e^{2V} \\
& \bar{W}_{\dot\alpha} = \frac{1}{8} D D e^{2V}\bar{D}_{\dot\alpha} e^{-2V}
\end{split}
\end{eqnarray}
which are chiral superfields since $\bar{D}_{\dot\alpha}W_{\beta}=D_{\alpha}\bar{W}_{\dot\beta}=0$.

The extended gauge transformation is  defined by 
\begin{eqnarray}
\begin{split}
 & e^{2V}  \,   \rightarrow  \, e^{2V^{\prime}} = e^{-i\bar\Lambda}e^{2V}e^{i\Lambda} \\
 & e^{-2V}  \, \rightarrow  \, e^{-2V^{\prime}} = e^{-i\Lambda}e^{-2V}e^{i\bar\Lambda},
\end{split}
\label{egs_app}
\end{eqnarray}
where
\begin{eqnarray}
 \bar{D}_{\dot\alpha}\Lambda = {D}_{\alpha}\bar\Lambda=0, 
\end{eqnarray}
with $\Lambda=\Lambda^a T^a$ and $\bar\Lambda=\bar\Lambda^a T^a$. 
The chiral superfields $W_\alpha$ and $\bar{W}_{\dot\alpha}$ transform in a covariant manner 
under the extended gauge transformation (\ref{egs_app}), 
\begin{eqnarray}
W_{\alpha} &\rightarrow& W_{\alpha}^{\prime} = e^{-i\Lambda}W_{\alpha}e^{i\Lambda},\\
\bar{W}_{\dot\alpha} &\rightarrow& \bar{W}_{\dot\alpha}^{\prime} = e^{-i\bar\Lambda}\bar{W}_{\dot\alpha}e^{i\bar\Lambda}. 
\end{eqnarray}
The infinitesimal extended gauge transformation is expressed as
\begin{eqnarray}
\delta V \aeq V^{\prime} -V \\
         \aeq
\frac{i}{2}{\cal L}_{V}\cdot
\left[ (\Lambda+\bar\Lambda) + \coth({\cal L}_{V})\cdot (\Lambda-\bar\Lambda)
\right],
\label{delta_V}
\end{eqnarray}
where the Lie derivative ${\cal L}_{V}\cdot \Lambda = [V,\Lambda]$. 

We can eliminates  $C, M, N, \eta, \bar\eta$ fields in (\ref{vector_superfield}) 
by choosing the component fields of $\Lambda, \bar \Lambda$ as the Wess-Zumino gauge fixing. 
The vector superfield in the Wess-Zumino gauge is given by  
\begin{eqnarray}
V_{\rm WZ}(x,\theta,\bar{\theta}) &=& -\theta\sigma_{\mu}\bar{\theta}A_{\mu}(x)
    +i\theta\theta\bar{\theta}\bar{\lambda}(x)
      -i\bar{\theta}\bar{\theta}\theta\lambda(x)
      \nonumber \\&& +\frac{i}{2}\theta\theta\bar{\theta}\bar{\theta}D(x).
\end{eqnarray}
Then, 
\begin{eqnarray}
\begin{split}
W_{\alpha}(y,\theta) &=
 -i\lambda_{\alpha}(y) + i\theta_\alpha D (y) - i (\sigma_{\mu\nu} \theta)_\alpha F_{\mu\nu}(y)
 \\&
+\theta\theta (\sigma_{\mu} D_{\mu}\bar\lambda)_{\alpha}(y) \\
\bar{W}^{\dot\alpha}(\bar{y},\bar\theta) &= 
i\bar\lambda^{\dot\alpha}(\bar y) + i \bar\theta^{\dot\alpha} D(\bar y) 
+i (\bar\sigma_{\mu\nu} \bar\theta)^{\dot\alpha} F_{\mu\nu}(\bar y)
\\& -\bar\theta\bar\theta (\bar\sigma_{\mu} D_{\mu}\lambda)^{\dot\alpha}(\bar y), 
\end{split}
\end{eqnarray}
where
\begin{eqnarray}
 D_{\mu}\lambda \aeq \partial_{\mu}\lambda +i[A_{\mu},\lambda],\\
 F_{\mu\nu} \aeq \partial_{\mu}A_{\nu}-\partial_{\nu}A_{\mu} +i[A_{\mu},A_{\nu}].
\end{eqnarray}

The ${\cal N}=1$ super Yang-Mills action in the Wess-Zumino gauge is given in terms of the component fields, 
\begin{eqnarray}
S_{\rm SYM} 
 \aeq     \frac{1}{g^2}\int d^4x\ {\rm tr}
     \left\{\frac{1}{2}F_{\mu\nu}^2+2i\bar{\lambda}\bar{\sigma}_{\mu}D_{\mu}\lambda+D^2\right\}(x).
     \nonumber\\
\end{eqnarray}
The action is invariant under supersymmetry transformation \cite{deWit:1975veh}, 
\begin{eqnarray}
\begin{split}
&
 \delta_{\xi} A_{\mu} = i\xi{\sigma}_{\mu}\bar{\lambda}
                         +i\bar{\xi}\bar{\sigma}_{\mu}\lambda \\
& \delta_{\xi} \lambda =  \sigma_{\mu\nu}\xi F_{\mu\nu} - \xi D \\
& \delta_{\xi} \bar\lambda =  \bar\sigma_{\mu\nu}\bar\xi F_{\mu\nu} + \bar\xi D \\ 
& \delta_{\xi} D       = i\xi\sigma_{\mu}D_{\mu}\bar{\lambda}
                          -i\bar{\xi}\bar{\sigma}_{\mu}D_{\mu}\lambda.    
\end{split}                       
\end{eqnarray}

\end{appendix}

%
\bibliographystyle{unsrt}
\bibliography{BibTex_template}

\end{document}